\documentclass[%
 reprint,
 superscriptaddress,
 amsmath,amssymb,
aps,
prb,
]{revtex4-1}

\usepackage{comment, color}
\usepackage{braket}

\usepackage{graphicx}
\usepackage{dcolumn}
\usepackage{bm}

\begin{document}

\title{Non-Hermitian band topology with generalized inversion symmetry}

\author{Ryo Okugawa}
 \affiliation{%
 	Graduate School of Information Sciences, Tohoku University, Sendai 980-8579, Japan
 }%
\author{Ryo Takahashi}
 \affiliation{%
	Department of Physics, Tokyo Institute of Technology, 2-12-1 Ookayama, Meguro-ku, Tokyo, 152-8551, Japan
}%

\author{Kazuki Yokomizo}
 \affiliation{%
 	Department of Physics, Tokyo Institute of Technology, 2-12-1 Ookayama, Meguro-ku, Tokyo, 152-8551, Japan
 }%
 \affiliation{%
	Condensed Matter Theory Laboratory, RIKEN, 2-1 Hirosawa, Wako, Saitama 351-0198, Japan
}%

\date{\today}

\begin{abstract}
Non-Hermitian skin effects and exceptional points are topological phenomena characterized by integer winding numbers.
In this study, we give methods to theoretically detect skin effects and exceptional points by generalizing inversion symmetry.
The generalization of inversion symmetry is unique to non-Hermitian systems.
We show that parities of the winding numbers can be determined from energy eigenvalues on the inversion-invariant momenta
when generalized inversion symmetry is present.
The simple expressions for the winding numbers allow us to easily analyze skin effects and exceptional points in non-Hermitian bands.
We also demonstrate the methods for (second-order) skin effects and exceptional points by using lattice models.
\end{abstract}

\maketitle

\section{Introduction}
Non-Hermitian physics has been recently investigated 
because the interplay of non-Hermiticity and topology creates many exotic phenomena beyond Hermitian physics \cite{Bergholtz21, Ashida20a}.
As examples, non-Hermitian skin effects and exceptional points have been intensively studied.
Their topological structures called point-gap topology \cite{Kawabata19X, Gong18, Zhou19} are unique to non-Hermitian systems.
Importantly, the non-Hermitian topological phenomena have been observed experimentally in various platforms
\cite{Zhen15, Zhou18, Cerjan19, Brandenbourger19, Hofmann20, Xiao20, Helbig20, Weidemann20, Chen20, Palacios20}. 

In non-Hermitian systems, energy spectra are strongly sensitive to boundary conditions.
The phenomenon is called a non-Hermitian skin effect \cite{Yao18L1}.
Skin effects give rise to states localized at the boundary under an open boundary condition (OBC),
which can be described by a non-Bloch band theory \cite{Yao18L1, Yao18L2, Lee19, Yokomizo19, Kawabata20, Yi20, Zhang20, Fu21}.
Because of the localization, 
analysis of skin modes is significant to non-Hermitian bulk-boundary correspondence
\cite{Yao18L1, Yao18L2, Kunst18, Yokomizo19, Liu19, Edvardsson19, Ezawa19B, Kunst19, Jin19, Lee19L, Okuma19, Imura19, Song19, Yang20, Yokomizo20P}.
Skin effects are also understandable as topological phenomena characterized by point-gap topology
\cite{Gong18, Ezawa19, Okuma20, Zhang20, Borgnia20, Okuma20a2, Denner20, Kawabata20a}.
Interestingly, recent works have revealed skin effects
in non-dissipative bosonic systems \cite{McDonald18, Yokomizo20a, Yang20a} and strongly correlated systems \cite{Okuma20a, Yoshida20a}.

As another topological object, exceptional points arise from band degeneracy between two states under a periodic boundary condition (PBC).
At exceptional points, eigenstates coalescence happens.
In general, band touching induces exceptional points \cite{Shen18, Yoshida18B, Kawabata19L, Yang21}
and lines \cite{Xu17, Cerjan18, Zyuzin18, Carlstrom18, Yang19, Wang19, Carlstrom19, Yang20L, Zhang20L}
in two- and three-dimensional non-Hermitian systems, respectively.
Intriguingly, exceptional points are topologically stable at generic points in the Brillouin zone (BZ).
As with skin effects, exceptional nodes have been investigated
in strongly correlated and disordered systems \cite{Yoshida18B, Yoshida19B, Zyuzin18, Moors19, Papaj19, Zyuzin19, Kimura19, Matsushita19, Michishita20, Nagai20}.

Symmetry also plays an important role to characterize band topology.
Symmetry classes are enriched in non-Hermitian systems
because transposition and complex conjugation for the Hamiltonians are inequivalent \cite{Kawabata19X}.
Accordingly, various symmetry-protected skin effects \cite{Okuma20, Rui19, Yoshida20, Liu20, Kawabata20B, Okugawa20}
and exceptional nodes \cite{Kawabata19L, Budich19, Okugawa19, Zhou19Opt, Yoshida19, Yoshida19B, Yokomizo20} have been theoretically suggested.
Meanwhile, crystal symmetries give constraints on band structures.
For instance, inversion symmetry prohibits a skin effect \cite{Kawabata19X, Liu19B, Yi20, Wu19}.
In Hermitian systems,
symmetry constraints have been utilized to diagnose band topology from occupied states only at the high-symmetry points
\cite{Fu07, Sato09, Sato10, Fang12, Hughes11, Turner12, Mondragon14, Kim15, Benalcazar17B, Kruthoff17, Po17, Ono18, Song18, Song18X, Tang19}.
Our purpose is to extend this idea to detect nontrivial non-Hermitian point-gap topology.

In fact, there are some difficulties in the computation for skin effects and exceptional points.
First, while a non-Bloch band theory can give details of skin modes,
it requires numerical precision to calculate energy eigenvalues in a large open system.
In addition, the non-Bloch band theory has not been completely established in high-dimensional multiband systems.
Second, because exceptional points and lines appear at generic points, we need to scan the entire BZ for the search.
Thus, we propose different methods based on crystal symmetry.

In this work, we study topological invariants that reveal whether skin modes and exceptional points appear by inversion symmetry.
Skin effects and exceptional points are characterized by winding numbers \cite{Okuma20, Kawabata19L}.
This study simplifies expressions of the winding numbers to reduce the calculation task by non-Hermitian symmetry. 
For the derivation,
we generalize inversion symmetry similarly to the ramification of nonspatial symmetries for non-Hermitian Hamiltonians.
We show that the winding numbers modulo 2 can be obtained from energy eigenvalues at the inversion-invariant momenta.
This analysis is beneficial to diagnose non-Hermitian band topology.

The paper is organized as follows.
In Sec.~\ref{WandGI}, we introduce generalized inversion symmetry for non-Hermitian Hamiltonians.
By generalized inversion symmetry,
we provide formulas to obtain the parity of the winding numbers for skin effects and exceptional points.
In Sec.~\ref{Modelskin}, we study skin effects in lattice models by using simplified expressions of the winding numbers.
Exceptional points and lines in lattice models are also analyzed with our topological invariants in Sec.~\ref{ModelEP}.
Our conclusion is summarized in Sec.~\ref{conclusion}.

\section{Winding numbers and symmetry} \label{WandGI}
\subsection{Topology for skin effects and exceptional points}
\begin{figure}[t]
	\includegraphics[width=8cm]{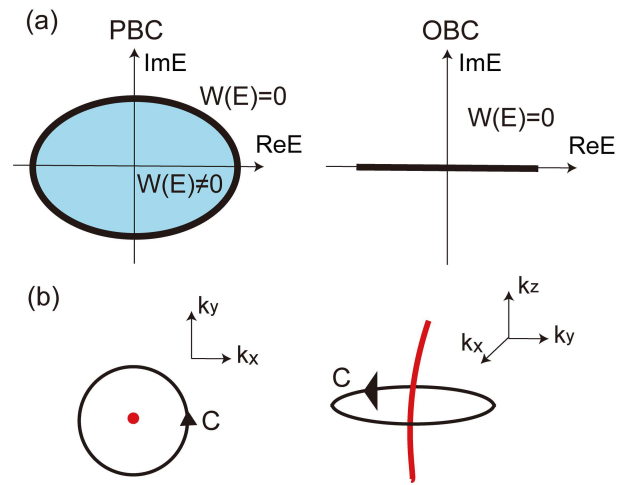}
	\caption{\label{Figwinding} (a) Complex energy spectra under the PBC and under the OBC.
		The thick lines represent the energy spectra closing the point gap.
		The winding number can be nonzero at each reference energy inside the energy spectrum under the PBC.
		As long as the point gap is open, the winding number does not change.
		(b) Topological characterization of an exceptional point and an exceptional line.
	}
\end{figure}

A non-Hermitian skin effect and an exceptional point are topologically characterized by winding numbers.
To see the topology,
we review point-gap topology for the topological characterization \cite{Gong18, Kawabata19X}.
A point gap for a Hamiltonian $H$ is open at a reference energy $E$ if $\det (H-E)\neq 0$.
When a point gap is open at $E$,
the topology for $H$ corresponds to that for the following extended Hermitian Hamiltonian given by
\begin{align}
	\tilde{H}=
	\begin{pmatrix}
		0 & H-E \\
		H^{\dagger}-E^* & 0
	\end{pmatrix}.
	\end{align}
Extended Hermitian Hamiltonians have additional chiral symmetry represented by $\Gamma \tilde{H} \Gamma ^{-1}=-\tilde{H}$ with
\begin{align}
	\Gamma = 
	\begin{pmatrix}
		1 & 0 \\
		0 & -1
	\end{pmatrix}.
	\end{align}
Therefore, $\tilde{H}$ belongs to a one-dimensional (1D) Hermitian class with chiral symmetry.
Because of the topological correspondence between $H$ and $\tilde{H}$,
$H$ under the PBC can be characterized by a winding number.
The integer winding number is \cite{Gong18, Kawabata19X, Okuma20}
\begin{align}
	W(E)=\int _{\mathrm{BZ}}\frac{dk}{2\pi i}\frac{d}{dk}\log \det [H(k)-E]. \label{windingskin}
	\end{align}
The integral is performed over the 1D first BZ.
Figure \ref{Figwinding} (a) shows a nontrivial winding structure of a complex spectrum under the PBC.
For point-gap topology, the winding number in Eq.~(\ref{windingskin}) reflects the topology of energy spectra rather than that of the eigenstates.

Essentially, no 1D non-Hermitian systems can have nontrivial winding structures of complex energy spectra under the OBC \cite{Okuma20, Zhang20}.
Therefore, energy spectra need to change the structures in the complex plane, depending on the boundary conditions.
The change leads to a non-Hermitian skin effect.
Hence, if a winding number is nonzero at a reference energy under the PBC,
a skin effect inevitably occurs under the OBC.
In terms of band topology,
the localization of skin modes originates from that of Hermitian zero-energy boundary modes through a nonzero winding number \cite{Okuma20}.

On the other hand, exceptional points appear from band touching
in two-dimensional (2D) non-Hermitian systems under the full PBC \cite{Shen18, Zhou18, Yoshida18B, Kawabata19L}.
Exceptional points close a point gap in the BZ.
Therefore, we can regard an exceptional point (EP) as a 1D topological phase transition in the 2D BZ.
Hence, an exceptional point at the energy $E_{\mathrm{EP}}$ is characterized by a 1D winding number to find the change of the topology.
To see the characterization, suppose that a 2D non-Hermitian system $H(\bm{k})$ has an exceptional point at energy $E_{\mathrm{EP}}$.
Then, the exceptional point can be characterized by the following winding number written as \cite{Kawabata19L}
\begin{align}
	W_{\mathrm{EP}}=\oint _C\frac{d\bm{k}}{2\pi i}\cdot \nabla _{\bm{k}}\log \det [H(\bm{k})-E_{\mathrm{EP}}], \label{windingEP}
\end{align}
where $C$ is a 1D integral path in the BZ.
A winding number in Eq.~(\ref{windingEP}) becomes nonzero if the path $C$ encircles the position of an exceptional point [Fig.~\ref{Figwinding} (b)].

Moreover, three-dimensional (3D) non-Hermitian systems can host an exceptional line from band touching between two states
since the point nodes become a line.
In a similar way to the 2D case, 
a winding number in Eq.~(\ref{windingEP}) characterizes exceptional lines.
A winding number is topologically nontrivial
if an exceptional line pierces any surface bounded by the integral path $C$.

\subsection{Generalized inversion symmetry}
We generalize an idea of crystal symmetry to grasp non-Hermitian band topology.
In general, extended Hermitian Hamiltonians can obtain symmetries that the original non-Hermitian Hamiltonians do not have \cite{Gong18, Kawabata19X}.
Therefore, additional crystal symmetries can also emerge for an extended Hermitian Hamiltonian, which affects topological characterization based on point-gap topology \cite{Kawabata20B, Okugawa20}.
Hence, we introduce generalized inversion symmetry
to give an extended Hermitian Hamiltonian with inversion symmetry, although the generalization is not unique \cite{Lee20, Kedia21}.

We define generalized inversion symmetry for a Hamiltonian $H(\bm{k})$ as
\begin{align}
	U_IH(\bm{k})U_I^{-1}=H^{\dagger}(-\bm{k}), \label{exinv}
	\end{align}
where $U_I$ is a unitary matrix that satisfies $U_I^2=1$.
When $H(\bm{k})$ is Hermitian, generalized inversion symmetry is just conventional inversion symmetry.
In the same way as conventional inversion symmetry,
we define inversion-invariant momenta as wavevectors satisfying $\bm{\Gamma} \equiv -\bm{\Gamma}$ modulo a reciprocal lattice vector.
Then, $H^{\dagger}(\bm{\Gamma})=U_IH(\bm{\Gamma})U_I^{-1}$ can be satisfied at the inversion-invariant momenta by choosing a proper gauge.
For example, inversion-invariant momenta in 3D systems are given by $\bm{\Gamma }_{i=(n_x,n_y,n_z)}=\sum _{j=x,y,z}n_j\bm{G}_j/2$,
where $n_{j=x,y,z}$ take the values $0$ and $1$, and $\bm{G}_{j=x,y,z}$ are three reciprocal lattice vectors.

Generalized inversion symmetry $U_I$ gives inversion symmetry to an extended Hermitian Hamiltonian $\tilde{H}$ with a real reference energy $E$.
When $E$ is real, $\tilde{H}$ acquires inversion symmetry $\tilde{I}$ represented by
\begin{align}
	\tilde{I}\tilde{H}(\bm{k})\tilde{I}^{-1}=\tilde{H}(-\bm{k}), ~~~
	\tilde{I}=
	\begin{pmatrix}
		0 & U_I \\
		U_I & 0
	\end{pmatrix},
	\end{align}
which satisfies  $\tilde{I}^2=\tilde{I}\tilde{I}^{\dagger}=1$.
The inversion operator $\tilde{I}$ anticommutes with the chiral operator $\Gamma$.
In other words, we have $\tilde{I}\Gamma =-\Gamma \tilde{I}$.
In this paper, we utilize inversion symmetry $\tilde{I}$ for topological characterizations of skin effects and exceptional points.
While we consider here a real reference energy,
generalized inversion symmetry can be defined for a complex reference energy.
We show the representation in Appendix \ref{GIS}.

We mention conventional inversion symmetry defined as $PH(\bm{k})P^{-1}=H(-\bm{k})$ for a non-Hermitian Hamiltonian $H(\bm{k})$.
Conventional inversion symmetry $P$ satisfies $P=P^{\dagger}$ and $P^2=1$.
In the presence of $P$, any extended Hermitian Hamiltonian has inversion symmetry
given by $\tilde{P}\tilde{H}(\bm{k})\tilde{P}^{-1}=\tilde{H}(-\bm{k})$ with $\tilde{P}=\mathrm{diag}(P,P)$.
Then, we have $\tilde{P}^2=\tilde{P}\tilde{P}^{\dagger}=1$, whereas $\tilde{P}\Gamma =\Gamma \tilde{P}$.
Generally, band topology depends on whether inversion and chiral operators commute or anticommute \cite{Chiu13, Morimoto13, Shiozaki14}.
When $\tilde{P}\Gamma =\Gamma \tilde{P}$, the 1D inversion-symmetric system does not show skin modes characterized by a winding number in Eq.~(\ref{windingskin}),
which stems from the topological difference.
In Appendix \ref{abs}, we discuss the absence of the skin effect due to conventional inversion symmetry.

\subsection{Parity of winding numbers for skin effects}
We derive simple formulas for a winding number in Eq.~(\ref{windingskin}) in the presence of generalized inversion symmetry.
To do so, we use the topological correspondence between non-Hermitian Hamiltonians and extended Hermitian Hamiltonians.
We set a reference energy $E$ to be real in order to exploit inversion symmetry hereafter.

\subsubsection{1D skin effect}
By using generalized inversion symmetry,
we study a 1D skin effect characterized by a winding number in Eq.~(\ref{windingskin}).
Before the discussion on the skin effect,
we introduce conventional topology for 1D Hermitian Hamiltonians with inversion and chiral symmetries to employ the topological correspondence.
If a 1D chiral-symmetric Hermitian Hamiltonian has inversion symmetry,
the parity of the winding number can be calculated from the number of states with negative parity eigenvalues below zero energy at the inversion-invariant momenta
\cite{Mondragon14, Hughes11, Turner12, Chiu16, Benalcazar17B}.
The expression is 
\begin{align}
	(-1)^W=(-1)^{n_-(0)-n_-(\pi )}, \label{windinginv}
\end{align}
where $n_-(0)$ and $n_-(\pi )$ are the number of states with negative parity eigenvalues below zero energy at $k=0$ and $\pi$, respectively.

As explained in Sec.~\ref{WandGI}, 
a winding number of a non-Hermitian Hamiltonian coincides with that of the extended Hermitian Hamiltonian.
Thus, if a non-Hermitian Hamiltonian has generalized inversion symmetry,
the parity of the winding number can be given from Eq.~(\ref{windinginv}) via topology in the extended Hermitian Hamiltonian.

We rewrite Eq.~(\ref{windinginv}) here for non-Hermitian cases.
We consider an $N\times N$ non-Hermitian Hamiltonian $H$ with generalized inversion symmetry $U_I$.
We assume that the point gap is open at a real reference energy $E$ under the PBC.
The extended Hermitian Hamiltonian at the inversion-invariant momentum $\Gamma _i$ is rewritten as
\begin{align}
	\tilde{H}(\Gamma _i)=
	\begin{pmatrix}
		0 & H(\Gamma _i)-E \\
		U_IH(\Gamma _i)U^{-1}_I-E & 0
	\end{pmatrix}.
	\end{align}
Since $U_IH(\Gamma _i)U_I^{-1}=H^{\dagger}(\Gamma _i)$, the matrix $U_I(H(\Gamma _i)-E)$ is Hermitian.
Hence, the eigenvalues $\lambda _n(\Gamma _i) ~(n=1, \dots ,N)$ are real.
We write the eigenvector with $\lambda _n (\Gamma _i)$ as $\ket{\lambda _n(\Gamma _i)}$.
The explicit eigenvalue equation is
\begin{align}
	U_I(H(\Gamma _i)-E)\ket{\lambda _n(\Gamma _i)}=\lambda _n (\Gamma _i)\ket{\lambda _n(\Gamma _i)}.
	\end{align}
All the eigenvalues $\lambda _n(\Gamma _i)$ are nonzero because $\det [U_I(H(\Gamma _i)-E)]\neq 0$ by assumption. 
As a result, we can find $2N$ eigenstates of $\tilde{H}(\Gamma _i)$
given by $\ket{p, p \lambda _n (\Gamma _i)}\equiv (U_I\ket{\lambda _n(\Gamma _i)}, p \ket{\lambda _n(\Gamma _i)})^T$ with $p=\pm 1$
because
\begin{align}	
	\tilde{H}(\Gamma _i)\left(
	\begin{array}{c}
		U_I\ket{\lambda _n(\Gamma _i)} \\ p\ket{\lambda _n(\Gamma _i)}
	\end{array}
	\right) = p \lambda _n (\Gamma _i) \left(
	\begin{array}{c}
		U_I\ket{\lambda _n(\Gamma _i)} \\ p \ket{\lambda _n(\Gamma _i)}
	\end{array} 
	\right) . \label{tildastates}
	\end{align}
Therefore, eigenstates of $\tilde{H}(\Gamma _i)$ can be constructed by those of $U_I(H(\Gamma _i)-E)$.
The states $\ket{p, p \lambda _n (\Gamma _i)}$ are also eigenstates of inversion symmetry $\tilde{I}$.
The eigenstates $\ket{+(-), +(-)\lambda _n (\Gamma _i)}$ have even (odd) parity because we have
\begin{align}
	\tilde{I}\ket{p, p \lambda _n (\Gamma _i)}=p \ket{p, p \lambda _n (\Gamma _i)}. \label{paritystates}
\end{align}
The simultaneous eigenstates can be also seen from the fact $\tilde{H}(\Gamma _i)$ and $\tilde{I}$ are block-diagonalized by the following unitary matrix:
\begin{align}
	Q=\frac{1}{\sqrt{2}}
	\begin{pmatrix}
		U_I & U_I \\
		1 & -1
	\end{pmatrix}.
\end{align}
By the unitary transformation $Q^{-1}\tilde{H}(\Gamma _i)Q$, $\tilde{H}(\Gamma _i)$ is block-diagonalized into the matrices $\pm U_I(H(\Gamma _i)-E)$.

\begin{figure}[t]
	\includegraphics[width=8cm]{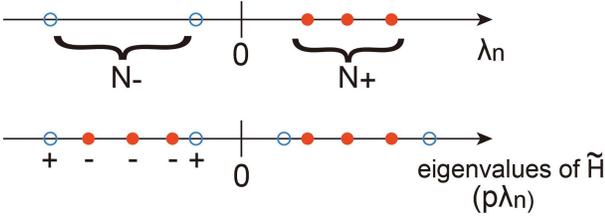}
	\caption{\label{specH} The relationship of eigenvalues of $U_I(H(\Gamma _i)-E)$ and $\tilde{H}(\Gamma _i)$ at the inversion-invariant momentum $\Gamma _i$
		on the real axes.
		The signs $\pm$ represent parity eigenvalues of the eigenstates of $\tilde{H}(\Gamma _i)$ below zero energy.
	}
\end{figure}

We can evaluate the parity of $W(E)$ for $H$ by using Eqs.~(\ref{tildastates}) and (\ref{paritystates}).
Let $N_{+(-)}(\Gamma _i)$ be the number of eigenvectors $\ket{\lambda _n(\Gamma _i)}$ with the positive (negative) eigenvalue at $\Gamma _i$ [Fig.~\ref{specH}].
By definition, $n_-(\Gamma _i)$ is the number of eigenstates $\ket{-, -\lambda _n(\Gamma _i)}$ below zero energy.
Therefore, the number $n_-(\Gamma _i)$ is equal to the number of corresponding eigenvectors $\ket{\lambda _n(\Gamma _i)}$ with positive $\lambda _n(\Gamma _i)$.
Namely, we have
\begin{align}
	n_-(\Gamma _i)=N_+(\Gamma _i)~(=N-N_{-}(\Gamma _i)). \label{Formula1}
	\end{align}
Equation (\ref{Formula1}) is one of the significant formulas to relate $\tilde{H}(\Gamma _i)$ to $H(\Gamma _i)$ in this paper,
and it holds true in any dimension.
In particular, when $U_I$ is the identity matrix,
$n_-(\Gamma _i)$ is directly connected with energy eigenvalues of $H(\Gamma _i)$.
Let $\varepsilon _n(\Gamma _i) ~(n=1, \dots ,N)$ be eigenvalues of $H(\Gamma _i)$.
Because we obtain $\lambda _n(\Gamma _i)=\varepsilon _n(\Gamma _i) -E$,
the number $n_-(\Gamma _i)$ becomes equal to the number of states whose eigenvalue satisfies $\varepsilon _n(\Gamma _i)>E$.

From Eqs.~(\ref{windinginv}) and (\ref{Formula1}), we can rewrite Eq.~(\ref{windingskin}) as
\begin{align}
	(-1)^{W(E)}&=(-1)^{N_+(0)-N_+(\pi)} \\
	&=(-1)^{N_-(0)-N_-(\pi)},
	\end{align}
where we have used $N=N_+(\Gamma _i)+N_-(\Gamma_i )$.
By using $(-1)^{N_-(\Gamma _i)}=\mathrm{sgn}(\det [U_I(H(\Gamma _i)-E)])$ and $(\det U_I)^2=1$,
we eventually obtain
\begin{align}
	(-1)^{W(E)}	=\prod _{\Gamma _i=0, \pi}\mathrm{sgn}(\det [H(\Gamma _i)-E]). \label{Formula2}
\end{align}
Now $\det [H(\Gamma _i)-E]$ takes the real values since $U_IH(\Gamma _i)U_I^{-1}=H^{\dagger}(\Gamma _i)$ and $E$ is real.
Equation (\ref{Formula2}) is a central result of this paper.
The formula allows us to easily calculate the parity of a 1D winding number for the skin effect.

We note that this formula in Eq.~(\ref{Formula2}) is also applicable
to diagnose (weak) topological skin effects in high-dimensional systems when only one direction is open.
For instance, we can apply the formula to a mirror skin effect in 2D mirror-symmetric systems \cite{Yoshida20}.
The mirror skin effect is characterized by the winding numbers of mirror sectors on the 1D mirror-invariant lines.
Therefore, if the mirror sectors have 1D generalized inversion symmetry, 
the parity of the winding number in each mirror sector is obtainable from Eq.~(\ref{Formula2}).

\subsubsection{2D second-order topological skin effect}
When a 2D non-Hermitian Hamiltonian has generalized inversion symmetry,
the system can show skin modes localized at the corner \cite{Okugawa20}.
The skin modes stem from intrinsic second-order topology protected by chiral and inversion symmetries.
The second-order topology can be also understood by a winding number
defined for a 1D ribbon geometry open in one direction.
While the winding number of a 1D ribbon describes zero-energy corner modes in the 2D Hermitian system \cite{Khalaf18B, Matsugatani18, Takahashi20},
it contributes to skin modes localized at the corner in the corresponding non-Hermitian system \cite{Okugawa20}.

To begin with, we introduce the 2D Hermitian second-order topology characterized by inversion symmetry.
Let us introduce a winding number $W_{x_j-\mathrm{OBC}}$ for an inversion-symmetric ribbon geometry open in the $x_j$ direction.
Because the ribbon geometry is periodic in the other direction, 
the winding number gives the number of topological zero-energy modes under the full OBC with the corners.
Importantly, if inversion symmetry is present,
the winding number can be evaluated
from the number of states with negative parity eigenvalues below zero energy at the inversion-invariant momenta in the 2D BZ.
The relationship is expressed as \cite{Okugawa20}
\begin{align}
	(-1)^{W_{x_j-\mathrm{OBC}}}=(-1)^{\mu _j/2}, 
\end{align}
where
\begin{align}
	\mu _{j=x,y}=n_-(0,0)-n_-(\pi ,\pi )+\tilde{s}_j[n_-(\pi, 0)-n_-(0, \pi )], \label{parityinv0}
	\end{align}
with $\tilde{s}_{x(y)} =+1(-1)$.
Here, $n_-(\bm{\Gamma} _i)$ is the number of states with negative parity eigenvalues below zero energy at the inversion-invariant momentum $\bm{\Gamma} _i$.
Therefore, when $\mu _{j=x,y} \equiv 2 \pmod 4$, it can lead to a nonzero winding number $W_{x_j-\mathrm{OBC}}$.
By the bulk-corner correspondence, we can have zero-energy corner modes under the full OBC if $W_{x_j-\mathrm{OBC}} \equiv 1 \pmod 2$.
We stress that $W_{x_j-\mathrm{OBC}}$ can be defined only when the ribbon geometry is gapped at zero energy.
For instance, if $\mu _j$ is an odd integer, the system necessarily has gapless points at zero energy \cite{Okugawa20, Ono18}.
Therefore, $W_{x_j-\mathrm{OBC}}$ cannot be defined if $\mu _j$ is an odd integer.

Henceforth, we investigate a second-order topological skin effect associated with inversion symmetry.
If a 2D non-Hermitian Hamiltonian $H$ has generalized inversion symmetry,
an extended Hermitian Hamiltonian $\tilde{H}$ with a real reference energy $E$ obtains inversion symmetry.
Then, $H$ and $\tilde{H}$ can be characterized by the same topological invariants $W_{x_j-\mathrm{OBC}}(E)$ and $\mu _j(E)$.
Because $W_{x_j-\mathrm{OBC}}(E)$ for $H$ is also defined for an inversion-symmetric ribbon open in the $x_j$ direction,
the winding number describes whether a skin effect occurs under the full OBC.
Because Eq.~(\ref{Formula1}) is valid in 2D systems, we have
\begin{align}
	\mu _{j}(E)=N_+(0,0)-N_+(\pi, \pi)+\tilde{s}_j[N_+(\pi ,0)-N_+(0, \pi )]. \label{parityinv}
	\end{align}
Therefore, if the point gap is open at $E$, we can obtain
\begin{align}
	&(-1)^{W_{x_j-\mathrm{OBC}}(E)} \notag \\
	&=(-1)^{\frac{1}{2}\{ N_{\pm}(0,0)-N_{\pm}(\pi, \pi)+\tilde{s}_j[N_{\pm}(\pi ,0)-N_{\pm}(0, \pi )]\} }.
	 \label{FormulaSOT}
	\end{align}
Thanks to generalized inversion symmetry,
we can evaluate the parity of $W_{x_j-\mathrm{OBC}}(E)$ by eigenvalues of $H$ under the full PBC.
Namely, if $\mu _{j}(E) \equiv 2 \pmod 4$ is obtained from Eq.~(\ref{parityinv}), 
we can easily detect second-order topological skin modes localized at the corner under the full OBC.

\subsection{Topological invariants for exceptional points}
In this section, we give a method to search for exceptional points in the BZ by generalized inversion symmetry.
Because band touching happens at any energy,
it is not easy to know energy $E_{\mathrm{EP}}$ of exceptional points.
Thus, a winding number cannot be used for the search in general.
Nevertheless, the energy can be typically set to be $E_{\mathrm{EP}}=0$ if the Hamiltonian has additional symmetry 
or if the energy origin can be theoretically shifted.
For example, two bands touch each other at zero energy 
when sublattice symmetry or parity-particle-hole symmetry is present. 
Therefore, we here assume that band touching closes a point gap at zero energy,
and discuss topological invariants for the exceptional points and lines.

\subsubsection{Exceptional points in 2D}
\begin{figure}[t]
	\includegraphics[width=8cm]{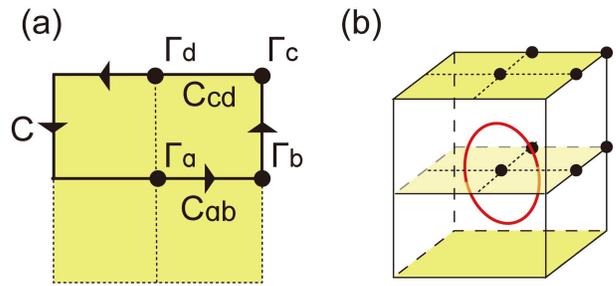}
	\caption{\label{FigEP} (a) The integral path in the winding number to detect exceptional points in the 2D BZ.
		The black points indicate inversion-invariant momenta in the BZ. 
		(b) An illustrative example of the exceptional line for ${\nu}_0^{\mathrm{EP}}\equiv 1 \pmod 2$.
	}
\end{figure}

We analyze exceptional points in 2D systems with generalized inversion symmetry.
Because $H^{\dagger}(-\bm{k})=U_IH(\bm{k})U_I^{-1}$, exceptional points appear at $\bm{k}$ and $-\bm{k}$ under the full PBC.
Thus, it is sufficient to search a half of the BZ.
For Eq.~(\ref{windingEP}), we choose an integral path $C$ which encircles a half of the 2D BZ, 
as illustrated in Fig.~\ref{FigEP}(a).
Here, we take a gauge to give $H(\bm{k})=H(\bm{k}+\bm{G}_i)$. 
Let $\nu ^{\mathrm{EP}}$ be the winding number with the path $C$.
Because of the gauge choice, the winding number $\nu ^{\mathrm{EP}}$ can be calculated from
\begin{align}
	\nu ^{\mathrm{EP}}=\left( \int _{C_{ab}}+\int _{C_{cd}}\right)
	\frac{d\bm{k}}{2\pi i}\cdot \nabla _{\bm{k}}\log \det H(\bm{k}).
	\end{align}
Here, $C_{ij}$ is the contour which contains the two inversion-invariant momenta $\Gamma _i$ and $\Gamma _j$,
as shown in Fig.~\ref{FigEP}(a).
In the same manner as the derivation of Eq.~(\ref{Formula2}),
the parity of the topological invariant $\nu ^{\mathrm{EP}}$ is obtained from
\begin{align}
	(-1)^{\nu ^{\mathrm{EP}}}=\prod _{\bm{\Gamma} _i}\mathrm{sgn}(\det H(\bm{\Gamma} _i)). \label{FormulaEP1}
	\end{align}
Therefore, when band touching happens at zero energy,
Eq.~(\ref{FormulaEP1}) reveals whether exceptional points exist in the 2D BZ.
Namely, if $\nu ^{\mathrm{EP}}\equiv 1 \pmod 2$, the system has exceptional points at $E_{\mathrm{EP}}=0$ in the BZ.

\subsubsection{Exceptional line in 3D}
Next, we generalize the 2D topological invariant in Eq.~(\ref{FormulaEP1}) to 3D non-Hermitian systems.
In 3D non-Hermitian systems, band touching can produce an exceptional line under the full PBC.
To find an exceptional line, 
we define the following topological invariant:
\begin{align}
	(-1)^{{\nu}_0^{\mathrm{EP}}}=\prod _{\bm{\Gamma} _i}\mathrm{sgn}(\det (H(\bm{\Gamma} _i)), \label{FormulaEP2}
	\end{align}
where the product is taken over all eight inversion-invariant momenta in the 3D BZ.
Below, we show that if ${\nu}_0^{\mathrm{EP}}\equiv 1 \pmod 2$,
an exceptional line lies at zero energy in the BZ.

Equation (\ref{FormulaEP2}) can be rewritten as
\begin{align}
	(-1)^{{\nu}_0^{\mathrm{EP}}}=(-1)^{\nu ^{\mathrm{EP}}(k_i=0)}(-1)^{\nu ^{\mathrm{EP}}(k_i=\pi )}, \label{sep4}
	\end{align}
where $k_i$ represents any of wave vectors $k_x, k_y$ and $k_z$. 
Equation (\ref{sep4}) means that ${\nu}_0^{\mathrm{EP}}$ can be described by a product of the invariants $\nu ^{\mathrm{EP}}$ for the two planes $k_i=0$ and $\pi$.
Thus, if ${\nu}_0^{\mathrm{EP}}\equiv 1 \pmod 2$, one of $\nu ^{\mathrm{EP}}(k_i=0 )$ and $\nu ^{\mathrm{EP}}(k_i=\pi )$ equals one modulo two.
Hence, the 3D system needs to possess at least one exceptional line which pierces one of the 2D planes $k_i=0$ and $k_i=\pi$,
as shown in Fig.~\ref{FigEP}(b).
As a consequence, if ${\nu}_0^{\mathrm{EP}}\equiv 1 \pmod 2$,
an exceptional line appears when band touching occurs at zero energy.

\section{Models for skin effects} \label{Modelskin}
We apply the formulas to investigate various skin effects in non-Hermitian lattice models with generalized inversion symmetry.
We evaluate the parity of winding numbers for a real reference energy $E$.
Whereas a reference energy is basically complex,
a winding number is unchanged as long as a point gap is open.
Therefore, we can focus on a winding number for a real reference energy $E$
to know whether a skin effect occurs.

\subsection{Generalized Hatano-Nelson model}
First, we study a 1D non-Hermitian one-band model.
The Hamiltonian is
\begin{align}
	H=\sum _{i}\sum _{m=1}^{M}[t_R^{(m)}c^{\dagger}_{i+m}c_i+t_L^{(m)}c^{\dagger}_ic_{i+m}],
	\end{align}
where $t_L^{(m)}$ and $t_R^{(m)}$ are real $m$-th nearest neighbor hopping parameters,
and the range is $M$.
When $M=1$, the Hamiltonian describes the Hatano-Nelson model without disorder \cite{Hatano96, Hatano97}.
Thus, this model is a generalization of the Hatano-Nelson model.

The model under the PBC is given by
\begin{align}
	h(k)=\sum _{m=1}^M[t_R^{(m)}e^{-ikm}+t_L^{(m)}e^{ikm}].
	\end{align}
The model has generalized inversion symmetry represented by $U_I=1$.
Therefore, we can apply Eq.~(\ref{Formula2}) to the model.
When a point gap is open at a real reference energy $E$,
the winding number modulo 2 can be calculated from
\begin{align}
	(-1)^{W(E)} 
		=&\mathrm{sgn}\left( \sum _{m=1}^M[t_L^{(m)}+t_R^{(m)}]-E \right) \notag \\
		\times &\mathrm{sgn}\left( \sum _{m=1}^M(-1)^m[t_L^{(m)}+t_R^{(m)}]-E \right). \label{onebandW}
	\end{align}
If the system has only the nearest-neighbor hopping, i.e. $M=1$,
the winding number is given by
\begin{align}
	(-1)^{W(E)}=\mathrm{sgn}(E-t_L^{(1)}-t_R^{(1)})\mathrm{sgn}(E+t_L^{(1)}+t_R^{(1)}).
	\end{align}
The Hatano-Nelson model shows the skin effect under the OBC
because the winding number is nonzero for $E<|t_L^{(1)}+t_R^{(1)}|$,
which agrees with previous works \cite{Okuma20, Zhang20}.

\begin{figure}[t]
	\includegraphics[width=8cm]{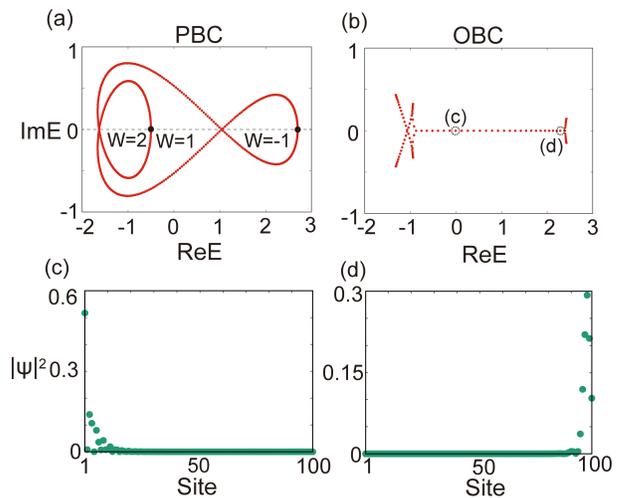}
	\caption{\label{oneband} (a)-(b) Energy eigenvalues for a one-band model under the PBC and under the OBC
		($t_L^{(1)}=1.0, t_R^{(1)}=0.8, t_L^{(2)}=0.6, t_R^{(2)}=0.5, t_L^{(3)}=-0.4$, and $t_R^{(3)}=0.2$).
		At the inversion-invariant momenta, the energy eigenvalues under the PBC are given by $h(0)=2.7$ and $h(\pi)=-0.5$, represented by the black points in (a).
		Despite the point-gap closing between $h(\pi)$ and $h(0)$,
		the parity of the winding number is unchanged between the two points on the real axis.
		(b) shows the skin effect associated with the winding number.
		(c)-(d) Spatial distribution $|\psi |^2$ of skin modes in the dotted circles in (b).
	}
\end{figure}

If a real reference energy $E$ lies between $h(0)$ and $h(\pi)$,
the winding number is always an odd integer when the band is gapped at $E$ . 
The case with $M=3$ is demonstrated in Fig.~\ref{oneband}(a).
As seen in Fig.~\ref{oneband} (b)-(d), skin modes necessarily appear under the OBC in this case.
Moreover, even though a point gap closes at a real reference energy between $h(0)$ and $h(\pi)$,
the parity of the winding number does not change because of constraints from generalized inversion symmetry.
Meanwhile, because a nonzero winding number signals a skin effect,
skin modes appear near the regions not only with $W(E)=\pm 1$ but also with $W(E)=2$.
Additionally, any winding number is invariant unless the point gap closes.
Hence, we can evaluate a winding number at a complex reference energy from the topology on the real energy axis.

\subsection{Non-Hermitian Su-Schriffer-Heeger model}
\begin{figure}[t]
	\includegraphics[width=8.0cm]{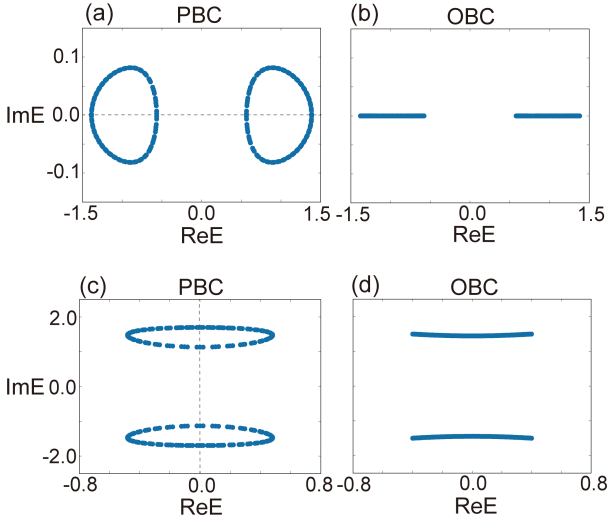}
	\caption{\label{NHSSH} Energy eigenvalues of the non-Hermitian SSH model.
		The spectra are calculated under the PBC in (a) and (c) and under the OBC in (b) and (d).
		We set $t_1=1.0, t_2=0.4 $ and $\delta =0.2$ for (a) and (b),
		and $t_1=1.0, t_2=0.4$ and $\delta =1.8$ for (c) and (d).
		In (a) and (c), the winding numbers under the PBC are nonzero on the real axis and on the imaginary axis inside the spectra, respectively.
	}
\end{figure}

As another representative example, we consider the Su-Schriffer-Heeger (SSH) model with a non-Hermitian term leading to a skin effect.
The non-Hermitian SSH model is described as \cite{Yao18L1, Yokomizo19}
\begin{align}
	H_{\mathrm{SSH}}(k)=(t_1+t_2\cos k)\sigma _x +t_2\sin k \sigma _y+i\delta \sigma _y
\end{align}
where $\sigma _x$ and $\sigma _y$ are Pauli matrices that represent two sublattices.
Here, $t_1$ and $t_2$ are real hopping parameters between the two sublattices, and $\delta$ is also a real parameter leading to non-Hermiticity.
The model has generalized inversion symmetry given by $U_I=\sigma _x$.
Thus, we can easily see that the skin effect happens by the winding number.
By using Eq.~(\ref{Formula2}), the parity of the winding number for a real reference energy $E$ can be calculated from
\begin{align}
	(-1)^{W(E)}=\prod _{s=\pm 1}\mathrm{sgn}[E^2+\delta ^2- (t_1+st_2)^2], \label{SSHwinding}
\end{align}
if the point gap is open at $E$.
As can be seen from Eq.~(\ref{SSHwinding}), $W(E)$ is nonzero on the real axis when the non-Hermiticity is weak.
Therefore, as the winding number is obviously nonzero from Fig.~\ref{NHSSH}(a),
skin modes collapse the point gap under the OBC [Fig.~\ref{NHSSH}(b)].

Interestingly, another inversion symmetry also emerges in the extended Hermitian Hamiltonian with a purely imaginary reference energy.
The corresponding generalized inversion symmetry is represented by $\sigma _yH_{\mathrm{SSH}}(k)\sigma _y=-H_\mathrm{{SSH}}^{\dagger}(-k)$ (see also appendix \ref{GIS}).
By the transformation $H(k) \rightarrow H'(k)\equiv -iH(k)$,
we can redefine generalized inversion symmetry for a real reference energy
as $\sigma _yH'_\mathrm{{SSH}}(k)\sigma _y = [H'_\mathrm{{SSH}}(-k)]^{\dagger}$ with the point-gap topology preserved.
Thus, a winding number for a purely imaginary energy $E'$ is also evaluated from 
\begin{align}
	(-1)^{W(E')}&=\prod _{\Gamma=0,\pi}\mathrm{sgn}\det [H'_{SSH}(\Gamma )+iE'] \\ 
	&=\prod _{s=\pm 1}\mathrm{sgn}[E'^2+\delta ^2- (t_1+st_2)^2].
	\end{align}
If $\delta $ is sufficiently larger than the hopping parameters, $W(E')$ can be finite on the imaginary axis [Fig.~\ref{NHSSH}(c)].
Then, skin modes also appear under the OBC, as shown in Fig.~\ref{NHSSH}(d). 

\subsection{Second-order skin effect}
We investigate a second-order skin effect in non-Hermitian systems with generalized inversion symmetry. 
We study the following 2D model given by
\begin{align}
	H(\bm{k})&=(m-c\sum_{j=x,y}\cos k_j)s_0 \notag \\
	&+it\sin k_ys_x+it\sin k_xs_y-B_xs_x-B_ys_y, \label{int}
	\end{align}
where the parameters $m, c, t, B_x$ and $B_y$ are real.
Here, $s_{x,y,z}$ are Pauli matrices and $s_0$ is the identity matrix.
Generalized inversion symmetry of the system is given by $U_I=s_0$, which can characterize a second-order topological skin effect.
Although Ref.~[\onlinecite{Okugawa20}] studied the second-order topological skin effect via the extended Hermitian Hamiltonian,
our method in this paper can directly diagnose it from the non-Hermitian Hamiltonian in Eq.~(\ref{int}).

\begin{figure}[t]
	\includegraphics[width=8cm]{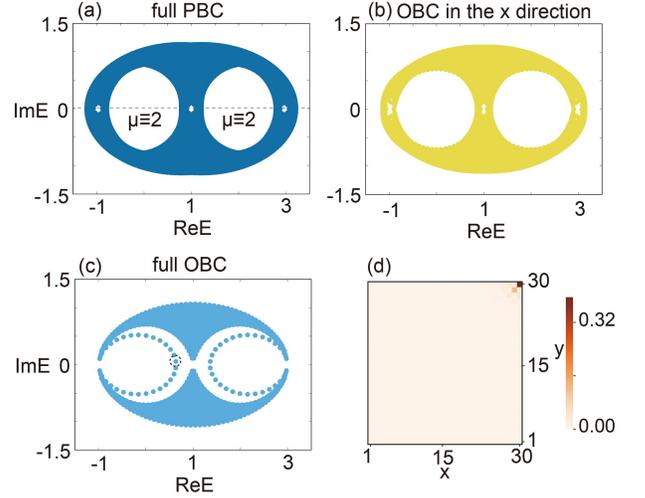}
	\caption{\label{SOskin} (a)-(c) Energy eigenvalues of the 2D non-Hermitian Hamiltonian under the PBC, the OBC only in the $x$ direction, and the full OBC. 
		We set $m=c=1.0, t=0.8$ and $B_x=B_y=0.15$ for the calculations.
		(c) shows second-order topological skin modes near the regions with $\mu (E)=-2\equiv 2 \pmod 4$.
		(d) Spatial distribution of the skin mode in the dotted circle in (c). The axes $x$ and $y$ indicate the coordinates.
	}
\end{figure}

Let us analyze the second-order skin effect by using parity invariants $\mu _j(E)$ in Eq.~(\ref{parityinv}).
Because $U_I$ is the identity matrix in this model,
$\mu _j(E)$ can be obtained from the energy eigenvalues at the the inversion-invariant momenta $\bm{\Gamma}_i$.
Namely, $N_{+}(\bm{\Gamma}_i)$ is given by the number of the eigenstates at $\bm{\Gamma }_i$
whose eigenvalue is larger than a real reference energy $E$.
The energy eigenvalues at the inversion-invariant momenta are
\begin{align}
	&\varepsilon _{\pm}(0,0)=m-2c \pm \sqrt{B_x^2+B_y^2}, \\
	&\varepsilon _{\pm}(0,\pi)=\varepsilon _{\pm}(\pi,0)=m\pm \sqrt{B_x^2+B_y^2}, \\
	&\varepsilon _{\pm}(\pi, \pi)=m+2c \pm \sqrt{B_x^2+B_y^2}.
\end{align}
Since $\varepsilon _{\pm}(0,\pi )=\varepsilon _{\pm}(\pi, 0)$,
we have $N_{\pm}(0,\pi )=N_{\pm}(\pi ,0)$.
Thus, the parity invariants $\mu_{x}$ and $\mu _{y}$ satisfy
\begin{align}
	\mu _{x}(E)=\mu _{y}(E)=N_{+}(0,0)-N_{+}(\pi, \pi ).
	\end{align}
For simplicity, we set $\mu (E)\equiv \mu _x(E)=\mu _y(E)$,
and we assume that $c>\sqrt{B_x^2+B_y^2}$.
Then, we have 
\begin{align}
\varepsilon _{-}(0,0)<\varepsilon _{+}(0,0)<\varepsilon _{-}(\pi, \pi)<\varepsilon _{+}(\pi, \pi).
\end{align}
Therefore, we can obtain $\mu (E)=-2$ for a real reference energy $E$ in the region $(\varepsilon _{+}(0,0), \varepsilon _{-}(\pi, \pi ))$ 
except $E=\varepsilon _{\pm}(0,\pi)$ on the real axis.
When the point gap at $E$ is open near the region with $\mu (E)=-2$,
the corresponding winding number $W_{x_j-\mathrm{OBC}}(E)$ is nonzero under the OBC only in one direction.
As a result, the system can show second-order topological skin modes under the full OBC.

We compute energy spectra under the various boundary conditions [Fig.~\ref{SOskin}(a)-(c)].
Within the regions with $\mu (E)=-2 \equiv 2 \pmod 4$,
the spectrum is gapped in $(\varepsilon _{+}(0,0), \varepsilon _{-}(0, \pi ))$ and $(\varepsilon _{+}(\pi ,0), \varepsilon _{-}(\pi , \pi ))$.
Figure.~\ref{SOskin} (c) shows skin modes around the regions with $\mu (E) \equiv 2 \pmod 4$.
The skin modes are indeed localized at the corner under the full OBC, as shown in Fig.~\ref{SOskin} (d).

\section{Models for exceptional points and lines} \label{ModelEP}
Finally, we confirm that the topological invariants in Eqs.~(\ref{FormulaEP1}) and (\ref{FormulaEP2}) can detect exceptional nodes in 2D and 3D lattice models.

\begin{figure*}[t]
	\includegraphics[width=17cm]{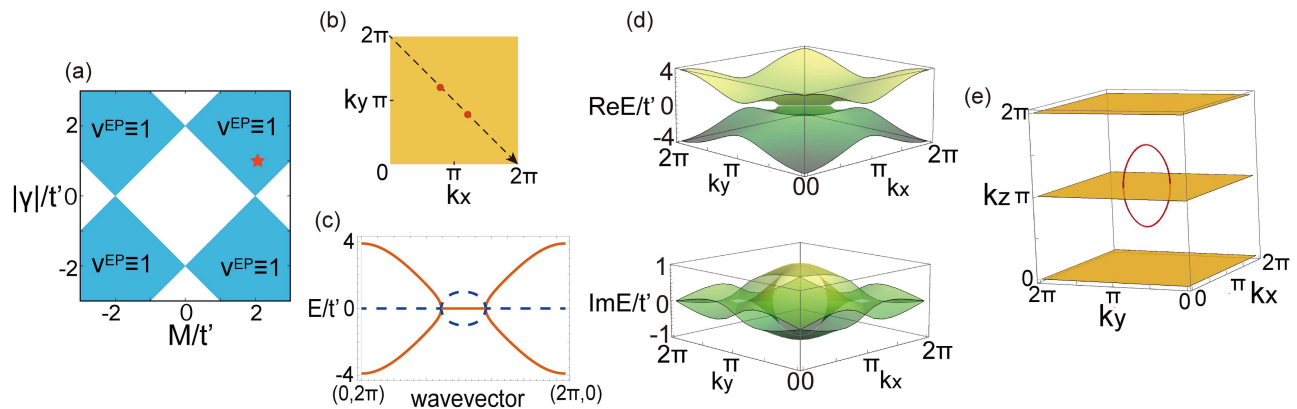}
	\caption{\label{phaseEP} (a) The parameter regions for the topological invariant $\nu ^{\mathrm{EP}}$.
		The blue-shaded regions give $\nu ^{\mathrm{EP}} \equiv 1 \pmod 2$.
		The star in the shaded region indicates the parameters ($M/t'=2.0$ and $|\gamma |/t'=1.0$) for the band calculation in the 2D model.
		(b),(c) The exceptional points in the 2D model with $M/t'=2.0, \gamma _x/t'=\gamma _y/t'=1/\sqrt{2}$ and $v_x/t'=v_y/t'=1.0$.
		The two points in (b) represent the positions of the exceptional points in the 2D BZ.
		In (c), the solid (dotted) lines are the real (imaginary) part of the energy bands along the dotted arrow in (b).
		(d) The real and the imaginary parts of energy bands in the 2D BZ.
		(e) The exceptional ring in the 3D model with $M/t'=3.0, t_z/t'=1.0, \gamma _x/t'=\gamma _y/t'=1/\sqrt{2}$ and $v_x/t'=v_y/t'=1.0$.
	}
\end{figure*}

\subsection{2D model with exceptional points}
We study a 2D non-Hermitian model that can have exceptional points.
The Hamiltonian under the full PBC is 
\begin{align}
	H_{\mathrm{2D}}(\bm{k})&=(M+t'\sum _{i=x,y}\cos k_i)\tau _x \notag \\
	&+(i\gamma _x+v_x\sin k_x)\tau _y+(i\gamma _y+v_y\sin k_y)\tau _z, \label{modelHEP}
	\end{align}
where $\tau _{x,y,z}$ are Pauli matrices.
All the parameters in the Hamiltonian are real.
The model has been also investigated as a non-Hermitian Chern insulator \cite{Yao18L2, Kawabata18B}.
The Hamiltonian has generalized inversion symmetry $U_I=\tau _x$.
Moreover, the Hamiltonian satisfies parity-particle-hole symmetry represented by
\begin{align}
	U_{CP}H^T_{\mathrm{2D}}(\bm{k})U_{CP}^{-1}=-H_{2D}(\bm{k}), ~U_{CP}=\tau _y.
	\end{align}
Thus, the band touching happens at zero energy.
Therefore, we can use Eq.~(\ref{FormulaEP1}) to search for exceptional points.

By Eq.~(\ref{FormulaEP1}),
the parity of the topological invariant $\nu ^{\mathrm{EP}}$ can be computed
from the eigenvalues at the inversion-invariant momenta $\bm{\Gamma}_{i=(n_x, n_y)}=(n_x\bm{G}_x+n_y\bm{G}_y)/2$.
In this model, we have
\begin{align}
	(-1)^{\nu ^{\mathrm{EP}}}=\prod _{n_{x,y}=0,1}&\mathrm{sgn}[|\gamma |+M+t'\sum _{i=x,y}(-1)^{n_i}] \notag \\
	&\times \mathrm{sgn}[|\gamma |-M-t'\sum _{i=x,y}(-1)^{n_i}],
	\end{align}
where we set $|\gamma |\equiv \sqrt{\gamma _x^2+\gamma _y^2}$.
Figure \ref{phaseEP} (a) shows parameter regions for $\nu ^{\mathrm{EP}}$ for the 2D model.

We search for exceptional points at zero energy in the 2D BZ.
When $\nu ^{\mathrm{EP}}\equiv 1 \pmod 2$, 
the 2D model necessarily has exceptional points at zero energy in the BZ.
In contrast, whether some parameter regions with $\nu ^{\mathrm{EP}} \equiv 0 \pmod 2$ have exceptional points depends on the other parameters $v_x$ and $v_y$.
Therefore, we calculate band structures of the model with $\nu ^{\mathrm{EP}} \equiv 1 \pmod 2$.
As shown in Fig.~\ref{phaseEP}(b)-(d), we can easily see that two exceptional points exist at zero energy in the model.

\subsection{3D model with an exceptional line}
We extend the model in Eq.~(\ref{modelHEP}) to a 3D system with generalized inversion symmetry.
We add a term $t_z \cos k_z \tau _x$ to the 2D Hamiltonian.
The 3D Hamiltonian is given by
\begin{align}
	H_{\mathrm{3D}}(\bm{k})&=(M+t'\sum _{i=x,y}\cos k_i +t_z \cos k_z)\tau _x \notag \\
	&+(i\gamma _x+v_x\sin k_x)\tau _y+(i\gamma _y+v_y\sin k_y)\tau _z.
	\end{align}
The 3D model also has generalized inversion symmetry and parity-particle-hole symmetry.

Since generalized inversion symmetry is present,
we can find an exceptional line by Eq.~(\ref{FormulaEP2}).
The topological invariant $\nu _0^{\mathrm{EP}}$ in this model is given by
\begin{align}
	(-1)^{\nu ^{\mathrm{EP}}_0}& \notag \\
	=\prod _{n_{x,y,z}=0,1}&\mathrm{sgn}[|\gamma |+M+t'\sum _{i=x,y}(-1)^{n_i}+t_z(-1)^{n_z}] \notag \\
	&\times \mathrm{sgn}[|\gamma |-M-t'\sum _{i=x,y}(-1)^{n_i}-t_z(-1)^{n_z}].
	\end{align}
If ${\nu ^{\mathrm{EP}}_0} \equiv 1 \pmod 2$, the system always has an exceptional line at zero energy.

We compute zero-energy eigenvalues of the 3D model with ${\nu ^{\mathrm{EP}}_0} \equiv 1 \pmod 2$ under the full PBC.
Figure \ref{phaseEP} (e) shows the exceptional ring in the 3D BZ.
The exceptional line pierces the $k_{i=x,y,z}=\pi$ planes.
The structure of the exceptional line is consistent with the discussion in Sec.~\ref{WandGI}.

\section{Conclusion and discussion} \label{conclusion}
In this paper, we have presented methods using topological invariants
to analyze point-gap topology in non-Hermitian systems with generalized inversion symmetry.
The simple formulas allow us to evaluate the parity of winding numbers for skin effects and exceptional points from energy eigenvalues only at inversion-invariant momenta.
The analysis is helpful to find skin effects and exceptional points even in multiband systems.
Our results can be applied to various skin effects such as a second-order skin effect and a mirror skin effect.
We have also given topological invariants to search for exceptional points and lines. 
The validity of our methods is confirmed by using lattice models.
 
Additionally, this work has clarified a relationship between point-gap topology and generalized inversion symmetry.
The generalized inversion symmetry is defined by using the symmetry ramification in non-Hermitian systems. 
The concept can be extended to other crystal symmetries.
Therefore, novel non-Hermitian band topology may be discovered by generalizing other crystal symmetries.
Moreover, a recent work has revealed that skin effects occur
when exceptional points and lines lie in the high-dimensional BZ \cite{Zhang21}.
Because our method can detect exceptional points,
the skin effects from exceptional nodes can be also predicted at the same time.

\textit{Note added}. We became aware of a related work which also studies symmetry indicators for non-Hermitian bands with pseudo-inversion symmetry included in generalized inversion symmetry \cite{Vecsei21}.

\begin{acknowledgments}
	This work was supported by JSPS Grant-in-Aid for Scientific Research on Innovative Areas “Discrete Geometric Analysis for Materials Design” Grant No. JP17H06469,
	and JSPS KAKENHI (Grants No. JP18J22113 and No. JP18J23289).
\end{acknowledgments}

\appendix
\section{Generalized inversion symmetry for complex reference points} \label{GIS}
In this appendix, we show that generalized inversion symmetry can be introduced to an extended Hermitian Hamiltonian with a complex reference energy $E$.
We define generalized inversion symmetry for a Hamiltonian $H(\bm{k})$ as
\begin{align}
	U_{I_\theta}H(\bm{k})U_{I_\theta}^{-1}=e^{2i\theta}H^{\dagger}(-\bm{k}), 	\label{UHU_theta}
\end{align}
where $U_{I_\theta}$ is a unitary matrix which satisfies ${U_{I_\theta}}^2=1$,
and $\theta$ is real. 
Equation (5) in the main text corresponds to the case of $\theta=0$ in Eq.~(\ref{UHU_theta}).  
When $\theta =\pi /2$, Eq.~(\ref{UHU_theta}) becomes $U_{I_\theta}H(\bm{k})U_{I_\theta}^{-1}=-H^{\dagger}(-\bm{k})$.
This case is also discussed for the non-Hermitian SSH model in Sec.~\ref{Modelskin}.
When $e^{-i\theta}E$ is real, $\tilde{H}(\bm{k})$ with $E$ has the following inversion symmetry $\tilde{I}_{\theta}$:
\begin{align}
	\tilde{I}_{\theta}\tilde{H}(\bm{k})\tilde{I}_{\theta}^{-1}=\tilde{H}(-\bm{k}), 
	\quad
	\tilde{I}_{\theta}=
	\begin{pmatrix}
		0 & e^{i\theta}U_{I_\theta} \\
		e^{-i\theta}U_{I_\theta} & 0
	\end{pmatrix}.
\end{align}
The inversion operator $\tilde{I}_{\theta}$ anticommutes with the chiral operator $\Gamma$, and $\tilde{I}_{\theta}^2=\tilde{I}_{\theta}\tilde{I}_{\theta}^{\dagger}=1$. 
Hence, generalized inversion symmetry can be defined even though reference points are complex.

We note that the generalized inversion symmetry for a complex reference energy is equivalent to that for a real reference energy in terms of non-Hermitian band topology. 
Let us consider the transformation $H(\bm{k})\to H_{\theta}(\bm{k}):=e^{-i\theta}H(\bm{k})$.
Then, we can rewrite Eq.~(\ref{UHU_theta}) as
\begin{align}
	U_{I_\theta}H_{\theta}(\bm{k})U_{I_\theta}^{-1}=[H_{\theta}(-\bm{k})]^{\dagger}. \label{Htheta}
\end{align}
Therefore, $U_{I_\theta}$ gives inversion symmetry to $\tilde{H_\theta}(\bm{k})$ with a real reference energy $E_{\theta}:=e^{-i\theta}E$. 
Because this transformation only rotates non-Hermitian bands on the complex plane,
the band topology does not change \cite{Gong18, Kawabata19N}.
Thus, we can evaluate a winding number $W(E)$ for $H(\bm{k})$ with $H_{\theta}(\bm{k})$.
For example, we give the formula corresponding to Eq.~(\ref{Formula2}).
For $e^{-i\theta}E \in \mathbb{R}$, we obtain
\begin{align}
	(-1)^{W(E)}
	&=\prod _{\Gamma _i=0,\pi}\mathrm{sgn}\det [H_{\theta}(\Gamma _i)-E_{\theta}] \\
	&=\prod _{\Gamma _i=0,\pi}\mathrm{sgn}\det [e^{-i\theta}(H(\Gamma _i)-E)].
\end{align}
As a result, various formulas in this paper can be used to see non-Hermitian band topology with generalized inversion symmetry in Eq.~(\ref{UHU_theta}).

\section{Absence of the skin effect by conventional inversion symmetry} \label{abs}
We show that conventional inversion symmetry $P$ prevents skin effects unless other symmetry protects them.
Although the non-Bloch theory can prove that conventional inversion symmetry prohibits skin effects \cite{Kawabata19X, Yi20},
we discuss it in view of the winding number in order to compare the two types of inversion symmetries.
When $PH(-k)P^{-1}=H(k)$ is satisfied,
we obtain
\begin{align}
	W(E)&=\int _{-\pi}^{\pi}\frac{dk}{2\pi i}\frac{d}{dk}\log \det [PH(-k)P^{-1}-E] \notag \\
	&=-\int _{-\pi}^{\pi}\frac{dk}{2\pi i}\frac{d}{dk}\log \det [H(k)-E]=0.
	\end{align}
Because the winding number for $H(k)$ becomes zero, the skin effect is not allowed to happen.

Here, we compare conventional inversion symmetry with generalized inversion symmetry.
Topological classification of Hamiltonians depends on (anti)commutation relation between crystal and chiral symmetries.
1D Hermitian inversion-symmetric systems in class AIII can become topologically nontrivial
only if the chiral operator anticommutes with the inversion operator \cite{Chiu13, Morimoto13}.
Therefore, our result is consistent with the topological classification.

\bibliographystyle{apsrev4-1}
\bibliography{skin-symm}

\end{document}